\documentclass[preprint,nofootinbib,12pt,floatfix,superscriptaddress,aps,prd,showpacs]{revtex4-1}
\usepackage[utf8]{inputenc} 
\usepackage{graphicx}
\usepackage{amsmath,amssymb,amsbsy,amsfonts,amsopn}
\usepackage{amstext}
\usepackage{color,xcolor}
\usepackage{slashed,esint}
\usepackage[dvips]{epsfig}
\usepackage[dvips]{graphicx}
\usepackage{units}
\usepackage{textcomp}
\usepackage{hyperref}
\newcommand{\beq}{\begin{equation}}
\newcommand{\eeq}{\end{equation}}
\newcommand{\bea}{\begin{eqnarray}}
\newcommand{\eea}{\end{eqnarray}}

\begin{document}

\title{Traversable wormholes from a smoothed string fluid in 4D Einstein--Gauss--Bonnet gravity}

\author{C. R. Muniz}
\email{celio.muniz@uece.br}
\affiliation{Universidade Estadual do Cear\'a, Faculdade de Educa\c c\~ao, Ci\^encias e Letras de Iguatu, 63500-000, Iguatu, CE, Brazil.}
\author{M. S. Cunha}\email{marcony.cunha@uece.br} \affiliation{Universidade Estadual do Cear\'a (UECE), Centro de  Ciências e Tecnologia (CCT), 60714-903, Fortaleza-CE, Brazil}
\author{L. C. N. Santos}\email{luis.santos@ufsc.br} \affiliation{Departamento de F\'isica, CFM - Universidade Federal de Santa Catarina; C.P. 476, CEP 88.040-900, Florian\'opolis, SC, Brasil}

\begin{abstract}
We investigate traversable wormhole solutions in four–dimensional Einstein–Gauss–Bonnet (EGB) gravity sourced by a smoothed string fluid. Originally proposed to model regular black holes, this energy density profile is adapted here to sustain wormhole geometries by allowing for a radially varying equation of state. We obtain zero–tidal–force solutions that satisfy all traversability criteria and remain globally regular. The Gauss–Bonnet (GB) coupling $\alpha$ plays a central role in shaping the throat geometry, and we identify a parameter region ($\alpha \geq 1$, $\varepsilon \leq 0.1$) in which the null energy condition is satisfied in the vicinity of the throat, representing a significant improvement over general relativistic counterparts. The interplay between the smoothing scale $a$ and the string density $\varepsilon$ ensures finite curvature invariants while reducing the violation of energy conditions. An analysis of the volume integral quantifier and the complexity factor further shows that strong EGB coupling simultaneously suppresses gravitational complexity and the total amount of exotic matter. These results establish a unified framework in which the same string fluid source can generate both regular black holes and stable traversable wormholes, depending on the strength of higher–curvature corrections.
\end{abstract}
\maketitle

\section{Introduction}

Dark energy and dark matter remain two of the most significant open problems in modern cosmology. While the cosmological constant and quintessence scalar fields have offered theoretical insights into cosmic acceleration \cite{Caldwell:1997ii, Peebles:2002gy}, alternative descriptions based on extended matter sources have gained prominence. In particular, string fluids -- characterized by an anisotropic energy-momentum tensor -- have been proposed as a unified framework to address issues ranging from the coincidence problem \cite{Kiselev:2002dx} to the flatness of spiral galaxy rotation curves \cite{Capozziello:2006ij, Soleng:1993yr, Sofue:2000jx}. Such fluids, often modeled with a tangential pressure equation of state (EoS) or as a \textit{cloud of strings} \cite{Vilenkin:1981kz,Nemiroff:2007xs,Arshad:2024qqj}, possess a profound impact on galactic scales and have been successfully applied to generate regular black hole solutions by smoothing the core curvature singularities \cite{NunesdosSantos:2025alw}.

In the realm of spacetime geometries, Lorentzian traversable wormholes \cite{Morris:1988cz} represent hypothetical shortcuts connecting distant regions of the universe. However, the stability and traversability of these structures within General Relativity (GR) typically demand the presence of exotic matter that violates the null energy condition (NEC) \cite{Visser:1995cc}. Given the successful application of string fluids and clouds of strings in modeling complex gravitational sources \cite{Richarte:2007bx,Sheikh:2023seo,Mustafa:2022obk,Gogoi:2022ove,Waseem:2023ejh}, they emerge as natural candidates to sustain such wormhole geometries, potentially minimizing the violation of energy conditions through their anisotropic nature.

While GR remains the standard paradigm for gravity in four dimensions -- validated by Lovelock's theorem \cite{Lovelock:1972vz} and observations such as gravitational waves \cite{LIGOScientific:2016aoc} and cosmic expansion \cite{Riess1998, Perlmutter1999} -- modified theories offer new avenues to address high-energy corrections. Einstein-Gauss-Bonnet (EGB) gravity \cite{Lovelock:1971yv, Boulware:1985wk}, a specific case of Lovelock theory, introduces higher-order curvature terms that are traditionally topological invariants in four dimensions. However, a recent regularization proposal by Glavan and Lin \cite{Glavan:2019inb} demonstrated that re-scaling the coupling constant allows for non-trivial EGB dynamics in $D=4$. Although this regularization initially faced scrutiny regarding its consistency and well-defined action \cite{LuPang2020,Gurses:2020ofy,Ai:2020peo,Shu:2020cjw,Mahapatra2020}, it was demonstrated that it is indeed possible to take the limit $D \rightarrow 4$ in the EGB action \cite{Hennigar:2020lsl, Fernandes:2020nbq}, generalizing a previous procedure used to take the limit $D \rightarrow 2$ in general relativity \cite{Mann:1992ar}. Since then, robust formulations and limiting procedures have since been established \cite{Gammon:2022bfu,Gammon:2023uss} and several wormhole solutions in $4D$ EGB gravity approach have
also been explored in the literature \cite{Jusufi:2020yus, Zhang:2020kxz, Yang2020WCCC, Liu:2021fzr, Mishra:2021qhi,  Oikonomou:2021kql, Godani:2024rqf, Hassan:2024xyx, Chakraborty:2025cpp, Muniz:2025teu} and, very recently, in \cite{DasPaul2026}.

Motivated by these developments, in this work we investigate traversable wormhole (TWH) solutions in $4D$ EGB gravity sourced by a \textit{smoothed string fluid}. This matter distribution was recently shown to generate regular black hole solutions in General Relativity when combined with a Dymnikova-type smoothing mechanism \cite{NunesdosSantos:2025alw}. Here, we demonstrate that the same energy density profile can also support TWH  geometries once the string fluid EoS is suitably relaxed. This allows us to establish a unified framework in which regular black holes (RBH) and TWH emerge from the same physical matter source, distinguished only by their geometric realization.

In light of these considerations, we perform a systematic analysis of traversable wormholes supported by a smoothed string fluid in $4D$ EGB gravity analyzing the geometric requirements for traversability, the behavior of curvature invariants, the linear equations of state of the anisotropic fluid, and the null energy condition. Particular attention is given to the role of the Gauss–Bonnet coupling and the string fluid parameter in controlling curvature regularity, energy condition violations, and the total amount of exotic matter, quantified through the volume integral measure. We also examine the complexity factor associated with the wormhole configuration, providing additional insight into the internal structure of the spacetime.

In this work, the supporting matter is modeled as a string fluid endowed with a radially varying equation of state, allowing a smooth interpolation between distinct physical regimes. Therefore, the paper is organized as follows: In Section II, we review the properties of a string fluid with a variable EoS  state. Section III presents the wormhole solutions and discusses their geometric and matter-source properties. Section IV analyzes the null energy condition and quantify the amount of exotic matter required to sustain the wormhole. Section V is devoted to the analysis of the complexity factor. Finally, in Section VI we summarize our results and discuss their physical implications.

\section{String fluid with a variable EoS}

To model matter sources capable of interpolating between vacuum–energy–dominated inner regions and string–dominated outer domains, we describe the matter content supporting the wormhole in terms of a string fluid endowed with a radially varying EoS. This framework is naturally formulated within the string cloud formalism originally introduced by Letelier \cite{letelier1979clouds}, in which a distribution of one–dimensional strings is described by a surface bivector $\Sigma_{\mu\nu}$ spanning the two–dimensional timelike worldsheet traced by the strings. The bivector is defined as
\begin{equation}
\Sigma^{\mu\nu}=\varepsilon^{AB}\frac{\partial x^{\mu}}{\partial \xi^A}\frac{\partial x^{\nu}}{\partial \xi^B},
\label{2.1}
\end{equation}
where $\varepsilon^{AB}$ is the antisymmetric Levi-Civita symbol in two dimensions, with $\varepsilon^{01} = -\varepsilon^{10} = -1$ and the parameters $\xi^A$ ($A = 0,1$), denote coordinates on the worldsheet, with $\xi^0$ being time-like and $\xi^1$ space-like.

The metric induced on the worldsheet by the spacetime metric $g_{\mu\nu}$ is given by 
\begin{equation}
h_{AB}= g_{\mu\nu}\frac{\partial x^{\mu}}{\partial \xi^A}\frac{\partial x^{\nu}}{\partial \xi^B}, \label{eq2.2}
\end{equation}
where the functions $x^\mu (\xi^A)$ describe the embedding of the string worldsheet in spacetime. Drawing an analogy with the energy–momentum tensor of a cloud of particles,
\begin{equation}
T^{\mu \nu} = \rho~u^{\mu} u^{\nu},
\label{2.3}
\end{equation}
in which $\rho$ denotes a energy density. 

Letelier proposed replacing the product $u^\mu u^\nu$ by the geometric object $\sqrt{-h}\frac{\Sigma^{\mu\lambda}\Sigma^{~\nu}_{\lambda}}{(-h)}$ \cite{letelier1979clouds}, which leads to the energy–momentum tensor of a cloud of strings
\begin{equation}
T^{\mu\nu}=\rho~\sqrt{-h}\frac{\Sigma^{\mu\lambda}\Sigma^{\:\:\nu}_{\lambda}}{(-h)},
\label{2.4}
\end{equation}
where $h$ refers to the determinant of the induced metric $h_{AB}$. 

This construction was later generalized to include pressure–like effects associated with string interactions \cite{letelier1981fluids}, leading to the extended energy–momentum tensor
\begin{equation}
T^{\mu\nu}=(p+\rho\sqrt{-h})\frac{\Sigma^{\mu\lambda}\Sigma^{\:\:\nu}_{\lambda}}{(-h)}+pg^{\mu\nu},
\label{2.5}
\end{equation}
where $p$ and $\rho$ denote, respectively, the pressure and density of the smoothed string fluid. For a static, spherically symmetric geometry, the symmetries of the spacetime restrict the bivector  $\Sigma_{\mu\nu}$ to have only two nonvanishing components, $\Sigma_{tr}$ and $\Sigma_{\theta\phi}$, while the determinant of the induced metric satisfies $h < 0$ \cite{Soleng:1993yr}. As a consequence, the energy–momentum tensor simplifies to
\begin{equation}
T_{t}^{t} = T_{r}^{r},  \qquad T_{\theta}^{\theta}=T_{\varphi}^{\varphi}= p.
\label{2.7}
\end{equation}

Following and generalizing the approach of Soleng \cite{Soleng:1993yr}, we allow the proportionality between density and transverse pressure to vary with the radial coordinate, i.e.,
\begin{equation}
\rho(r) = \bar{\alpha}(r) p(r),
\end{equation}
which leads to the energy–momentum tensor \cite{NunesdosSantos:2025alw}
\begin{equation}
T^{\mu}_{~\nu}=\left[- \rho(r),- \rho(r),\frac{\rho(r)}{\bar{\alpha}(r)},\frac{\rho(r)}{\bar{\alpha}(r)}\right].
\label{2.8}
\end{equation}
The limit $\bar{\alpha} \rightarrow \infty$ corresponds to vanishing transverse pressure ($p \rightarrow 0$) and therefore recovers the case of a pure cloud of strings. 

In the context of traversable wormhole solutions explored in this work, The radial dependence of $\bar{\alpha}(r)$ encodes a variable transverse EoS, which allows the string fluid to interpolate between different physical regimes. In particular, we identify
\begin{equation}
    \bar{\alpha}(r) = \frac{1}{\omega_t(r)},  
\end{equation}  
where $\omega_t(r)$ is the transverse linear equation-of-state parameter. This choice makes explicit how the smoothed string fluid can mimic, in different radial domains, vacuum–like behavior or cosmic–string–like matter. 

Although the energy–momentum tensor enforces $p_r = - \rho$, this condition is reinterpreted here as a de Sitter–like radial equation of state rather than as a black–hole horizon constraint. In the context of traversable wormhole geometries, we instead require
\begin{equation}  
    \omega_r(r) \to -1,
\end{equation}  
so that the radial pressure approaches a de Sitter–like form near the throat while the transverse sector controls the string–dominated behavior at larger radii (see below that $\omega_r(r)$ is the radial linear EoS parameter).

\section{Wormhole solutions: Geometry and source properties}\label{wsolutions}

The smoothed string fluid endowed with a radially varying equation of state introduced in previous section admits a natural reinterpretation in terms of an effective Kiselev–type anisotropic matter source. In this correspondence, the transverse equation–of–state parameter $\omega_t(r)$ of the string fluid plays the role of the Kiselev parameter $\omega_q$, allowing the matter distribution to interpolate smoothly between a de Sitter–like regime in the inner region and a cosmic–string–dominated behavior at large radii. 


In this section, we construct traversable wormhole solutions in four-dimensional EGB gravity sourced by a smoothed string fluid. We analyze both the geometric requirements for traversability and the physical properties of the corresponding matter source. Throughout this section, we focus on zero-tidal-force configurations, which provide the simplest realization of traversable wormholes.

\subsection{The Morris-Thorne metric in the EGB theory}
The Morris–Thorne metric describing a static, spherically symmetric traversable wormhole given by \cite{Morris:1988cz}
\begin{equation}\label{morris}
    ds^2=-e^{2\Phi(r)}dt^2+\frac{dr^2}{1-\frac{b(r)}{r}}+r^2d\Omega^2,
\end{equation}
where $\Phi(r)$ and $b(r)$ are functions of the radial coordinate $r$, known as the redshift function and the shape function, respectively. The throat of the wormhole is located at $r=r_0$, where the boundary condition $b(r_0) = r_0$ must be satisfied when radial coordinate decreases from the spacial infinity to this (minimum) value. The quantity $d\Omega$ denotes the line element on a unitary $2$--sphere.

As discussed in the introduction, while the $D \to 4$ limit has been subject to scrutiny, the field equations to be considered here are consistent with regularized 4D-EGB theories \cite{LuPang2020, Hennigar:2020lsl}. In this context, for the metric \eqref{morris} above, the four-dimensional EGB field equations yield the energy density and pressures in the form \cite{Farooq:2023rsp,Chakraborty:2025cpp,Cunha:2025jzh}
\beq \label{eqrho}
8\pi \rho(r)
= \frac{b(r)}{r^3} \left( 1 - \frac{\alpha b(r)}{r^3} \right) 
+ \frac{1}{r^2} \left(1 + \frac{2\alpha b(r)}{r^3} \right) 
\left(b'(r) - \frac{b(r)}{r} \right)
\eeq
\begin{equation}\label{radpress}
8\pi p_r(r) 
= -\frac{b(r)}{r^3} \left( 1 - \frac{\alpha b(r)}{r^3} \right) 
+ \frac{2}{r} \left( 1 - \frac{b(r)}{r} \right) 
\left(1+ \frac{2\alpha b(r)}{r^3} \right) \Phi'(r)
\end{equation}
\bea\label{transpress}
8\pi p_t(r)  
&=& -\frac{\alpha b(r)^2}{r^6} + \frac{2 \alpha}{r^4} \left(1 - \frac{b(r)}{r} \right) \left( b(r) - r b'(r) \right) \Phi'(r)\nonumber\\
&+& \left(1 - \frac{b(r)}{r} \right) \left[ 
\left(1 - \frac{2 \alpha b(r)}{r^3} \right)
\left( \frac{b(r) - r b'(r)}{2r^2(r - b(r))} - \frac{\Phi'(r)}{r} \right)\right.\\
&+& \left. \left(1+\frac{2 \alpha b(r)}{r^3} \right)
\left( \frac{(b(r) - r b'(r)) \Phi'(r)}{2r(r - b(r))} + \Phi''(r) - \Phi'(r)^2 \right)\right],\nonumber
\eea
where primes denote derivatives with respect to $r$. As in the General Relativity limit, the energy density determines the shape function through an effective mass function $m(r)$ defined by 
\beq
m(r) = \int  4 \pi r^2 \rho(r) dr.  \label{FuncionM} \\
\eeq
Besides this, notice that equation (\ref{eqrho}) can be written as  
\bea
   && 2 \cdot 4 \pi r^2 \rho (r)= b'(r)+ \alpha \left ( \frac{\left (b(r)/r \right )^2}{r}  \right )'. \nonumber
\eea
Integrating, we have
\bea
   && 2 \cdot m(r) = b(r) + \frac{\alpha}{r^3} b(r)^2.
\eea  
Solving the resulting quadratic equation for $b(r)$, one obtains two branches. Requiring a smooth General Relativity limit as $\alpha \rightarrow 0$ selects the physically admissible branch, leading to 
\begin{equation}
b(r)= \frac{r^3}{2\alpha} \left(-1 + \sqrt{1 + \frac{8 \alpha}{r^3}m(r)}\right), \label{eqparaB}  
\end{equation}
which guarantees that the radial metric component,
\begin{eqnarray}
 g_{rr}^{-1}=1-\frac{b(r)}{r} = 1 - \frac{r^2}{2 \alpha} \left (- 1 + \sqrt{1 + \frac{8  \alpha}{r^3}m(r)}\right), \label{grr}  
\end{eqnarray}
reduces continuously to its General Relativity counterpart when the Gauss–Bonnet coupling is switched off, that is,
%
\begin{eqnarray}
    \displaystyle \lim_{\alpha \to 0} g_{rr}^{-1} = 1 - \frac{2m(r)}{r}. \label{LimiteGR}
   \end{eqnarray}

\subsection{Smoothed string fluid source and shape function}
\begin{figure}[hb!]
 \centering
    \includegraphics[width=0.65\linewidth]{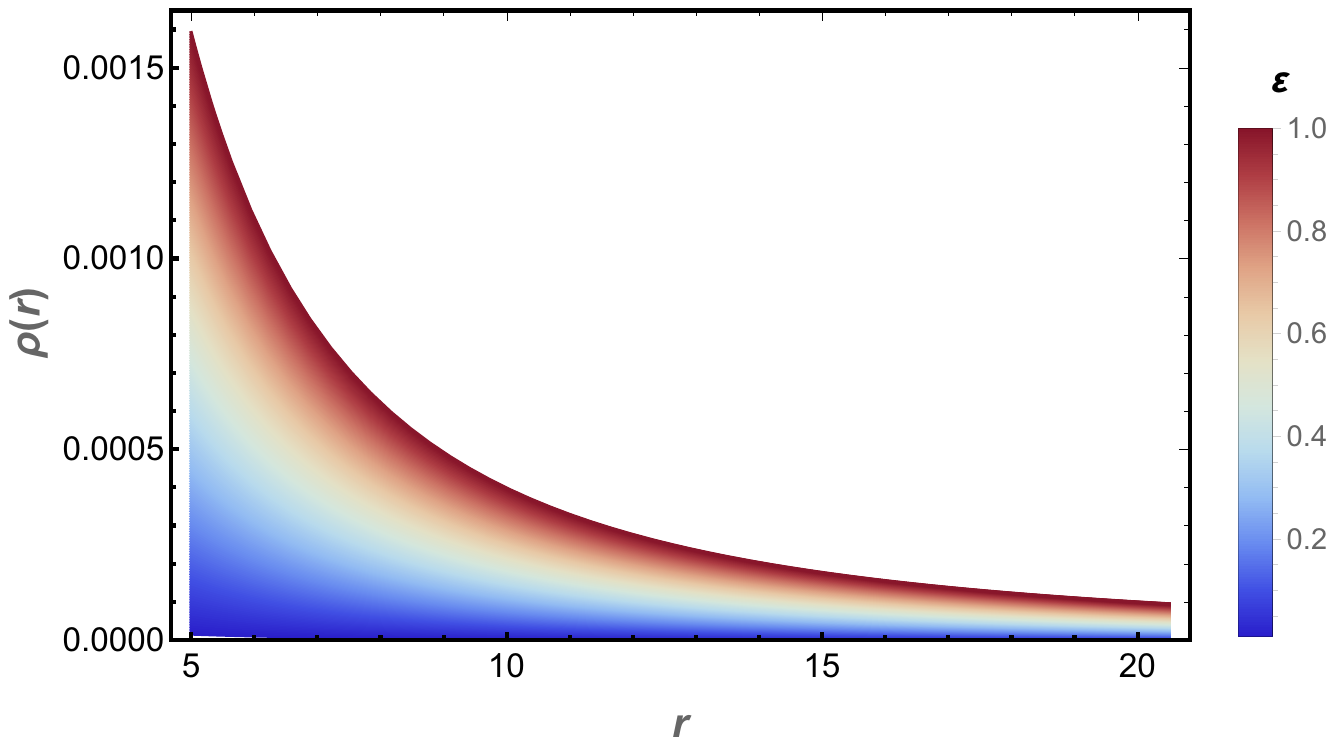}
    \caption{Color gradient for the energy density $\rho(r)$ for $\varepsilon=[0,1]$, with $a=1$ and $r_0=5$.}
    \label{fig_rho}
\end{figure}

Motivated by reference \cite{NunesdosSantos:2025alw}, we adopt the following smoothed string fluid energy density, 
\begin{equation}
\rho(r)=\frac{
\varepsilon a^3 e^{\frac{r^3}{a^3}} + \left(3 \varepsilon r^3 - \varepsilon a^3 + 3 r_g r^2\right)}
{8\pi a^3 e^{\frac{r^3}{a^3}} r^2}, \label{rho}
\end{equation}
where $a$ sets the regularization scale and $\varepsilon$ controls the amount of string fluid in the spacetime. The Schwarzschild radius $r_g$ is identified with the throat radius $r_0$. We illustrate the behavior of the energy density with $r$ and $\varepsilon$ in a color gradient in figure \ref{fig_rho}.

This energy density was previously shown to generate a regular black hole solutions supported by an anisotropic string fluid \cite{NunesdosSantos:2025alw}. Here, we take a natural yet nontrivial step further by investigating the same matter distribution in the context of EGB modified gravity, identifying it as a potential source for TWH geometries once the fluid EoS is relaxed. This extension is physically meaningful, since both regular black holes and wormholes require the absence of curvature singularities.

Integrating equation \eqref{FuncionM} with the energy density \eqref{rho}, we obtain 
\begin{equation}
    m(r)= \frac{1}{2} \left[\beta+ \varepsilon r -(\varepsilon r+r_0) \exp(-r^3/a^3)  \right].
\end{equation} 
\begin{figure}[hb!]
   \includegraphics[width=0.495\textwidth]{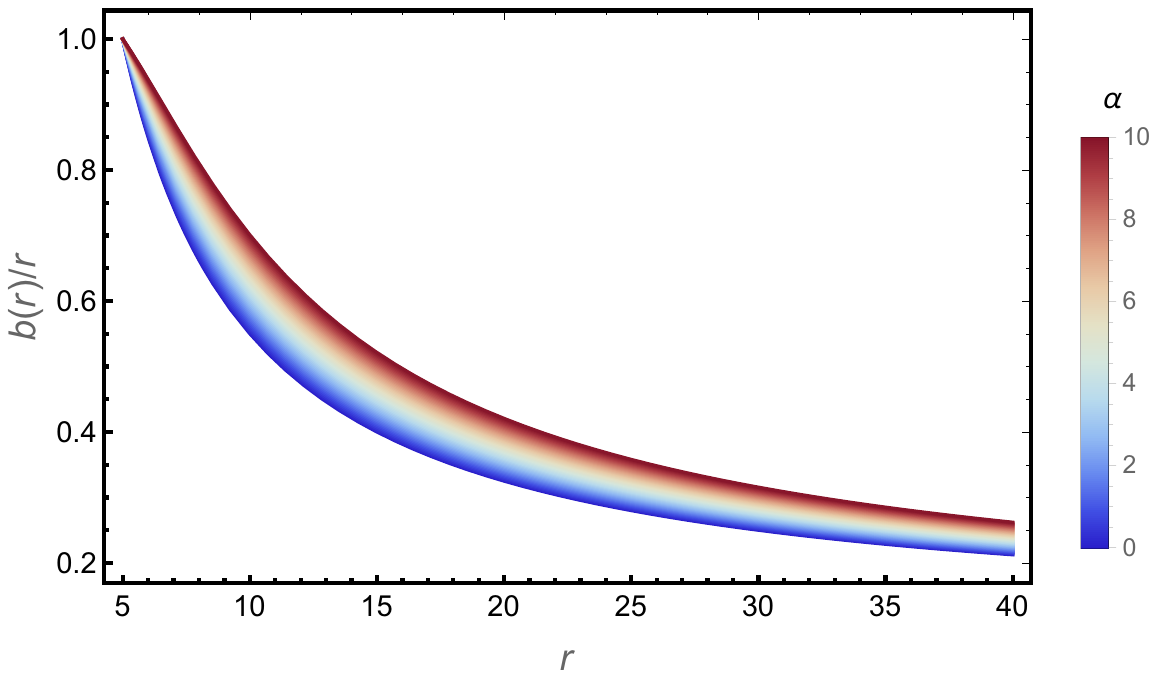}	
   \includegraphics[width=0.482\textwidth]{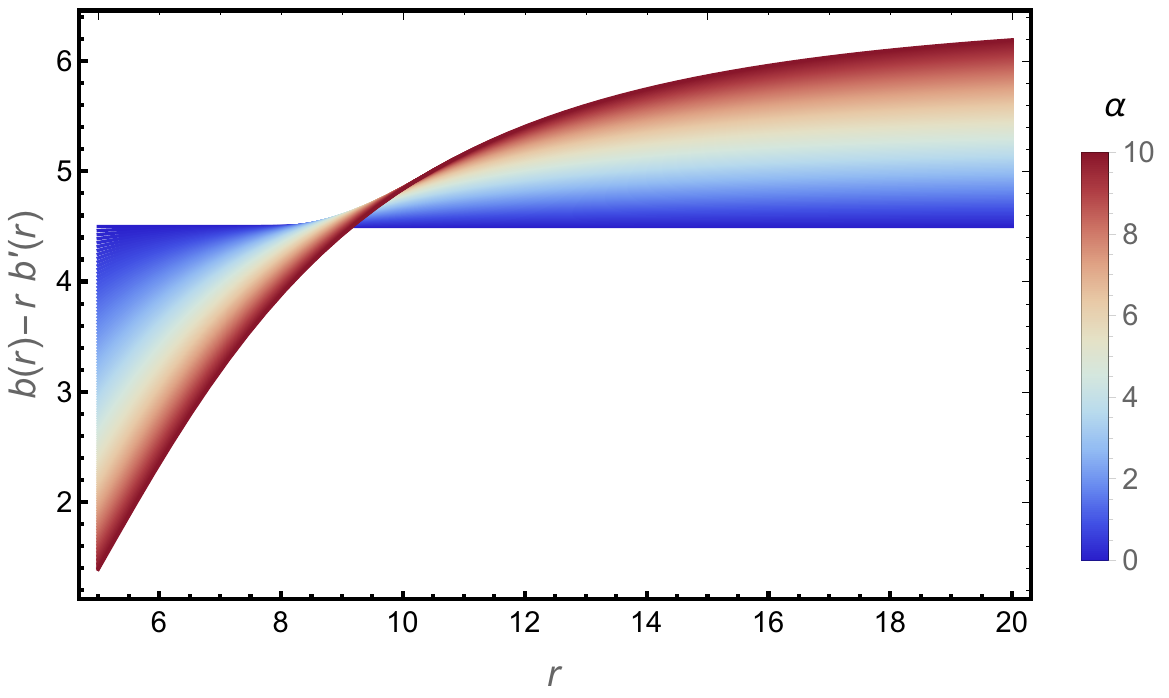}	
   \includegraphics[width=0.495\textwidth]{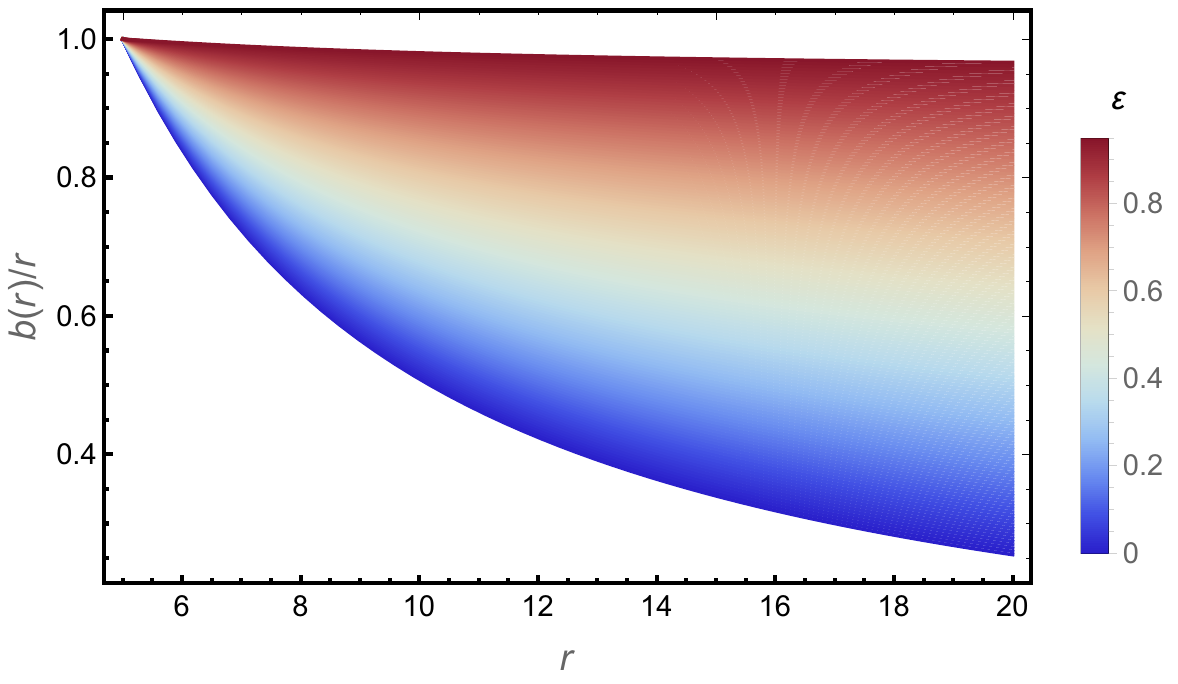}
   \includegraphics[width=0.482\textwidth]{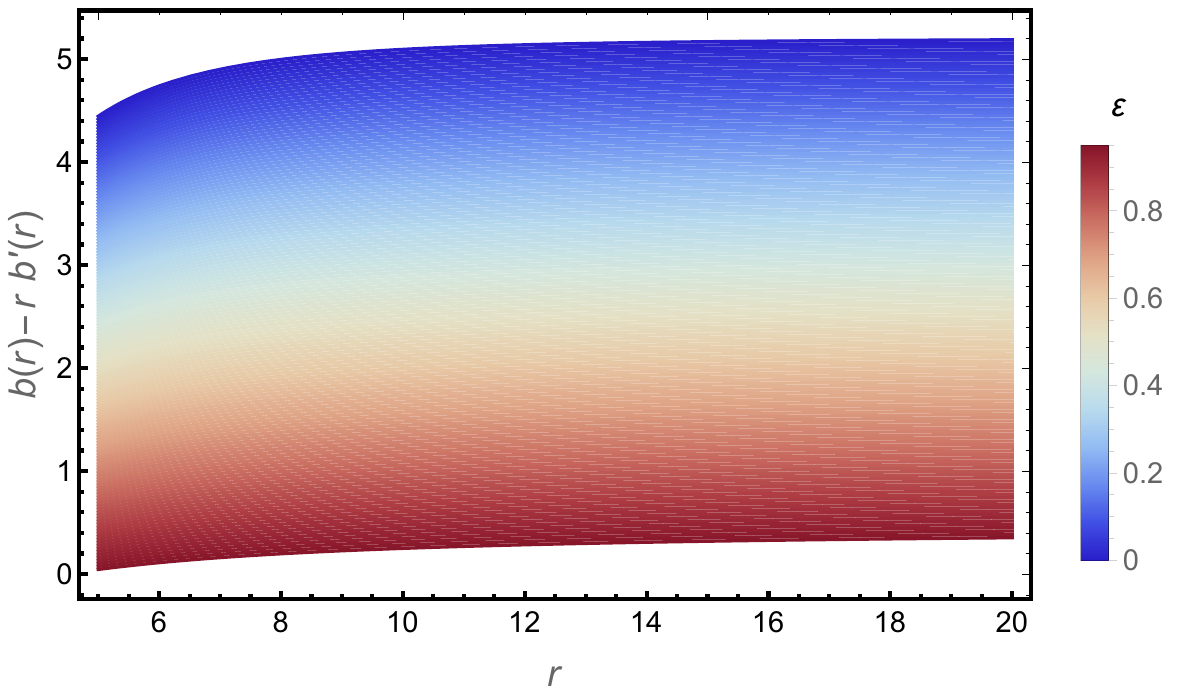}	
\caption{Top left panel: The geometric condition (2) $b(r)/r$, with $r_0 = 5$, $a=1$, and $\varepsilon = 0.1$.  Top right panel: The geometric condition (4) for $b(r)-r b'(r)$ (flaring-out), considering the same parameters. Bottom left panel: Dependence of $b(r)/r$ on $\varepsilon$, with $r_0=5$, $a = 1$, and $\alpha=0.1$ fixed. Bottom right: Flaring-out condition $b(r) - r b'(r)$ for the same $\varepsilon$-dependent case. In the figures, the color gradient represents  $\alpha \in [0,10]$ (top) and $\varepsilon \in [0,1]$ (bottom), as shown in the vertical colorbars, respectively.}
    \label{all_conditions}
\end{figure}
which leads to the shape function
  \begin{equation}\label{b(r)}
    b(r)= \frac{r^3}{2 \alpha} \left [ -1+ \sqrt{1+\frac{4\alpha}{r^3} \Big ( \beta + \varepsilon r - \left ( \varepsilon r + r_0 \right ) \exp(-r^3/a^3)\Big )} \right].
\end{equation}
Imposing the throat condition $b(r_0)= r_0$ fixes the integration constant $\beta$, that is,
\begin{equation} \label{ConstanteBeta}
    \beta =r_0 (1+\varepsilon)\exp (-r_0^3/a^3) + r_0 (1-\varepsilon) + \frac{\alpha}{r_0}.
\end{equation}
The asymptotic behavior $b(r)/r \rightarrow 0$ as $r\rightarrow \infty$ follows directly, ensuring asymptotic flatness. 

\subsection{Geometric conditions for traversability}
    Traversable wormholes must satisfy several geometric conditions \cite{Morris:1988cz, Visser:1995cc}, among them:
\begin{enumerate}
    \item  \label{Condicion1} The throat condition $b(r_0)=r_0$
    
    \item \label{Condicion2} The proper radial distance
    \begin{equation}
    l(r) = \pm \int_{r_0}^{r} \frac{dr}{\sqrt{1 - \frac{b(r)}{r}}},
    \end{equation}
    must be a real number. For this, it is necessary that
    \begin{equation}
      1 - \frac{b(r)}{r} \geq 0,  
    \end{equation}
    for all values of $r \geq r_0$.
    
    \item \label{Condicion3} The asymptotic flatness condition requires that
   \begin{equation}
    \lim_{r \to \infty} \frac{b(r)}{r} \to 0.
   \end{equation}
    This ensures that at infinity, the proper radial distance $l \to \pm \infty$.

    \item \label{Condicion4} The flaring-out condition, ${b(r) - b'(r)r} > 0$, for all values of $r \geq r_0$, ensuring the outward expansion of the TWH throat.

     \item \label{Condicion5} The absence of horizons, requiring $\Phi(r)$ to be finite for all values of $r \geq r_0$.
    
\end{enumerate}

We focus on zero-tidal-force wormholes by choosing $\Phi=\text{const}$, which automatically eliminates horizons. Therefore, conditions (i), (iii), and (v) are satisfied by construction. The remaining conditions are verified numerically and illustrated in figure \ref{all_conditions}, that is, the results show that both the positivity of $0<b(r)/r<1$ (top left and bottom left panels) and the flaring-out condition (top right and bottom right panels) are satisfied for wide ranges of the parameters $\alpha$ and $\varepsilon$. Increasing the Gauss–Bonnet coupling smooths the geometry near the throat, while the parameter $\varepsilon$ controls the overall strength of the string fluid contribution. 

\begin{figure}[hb!]
    \centering 
    \includegraphics[width=0.50\textwidth]{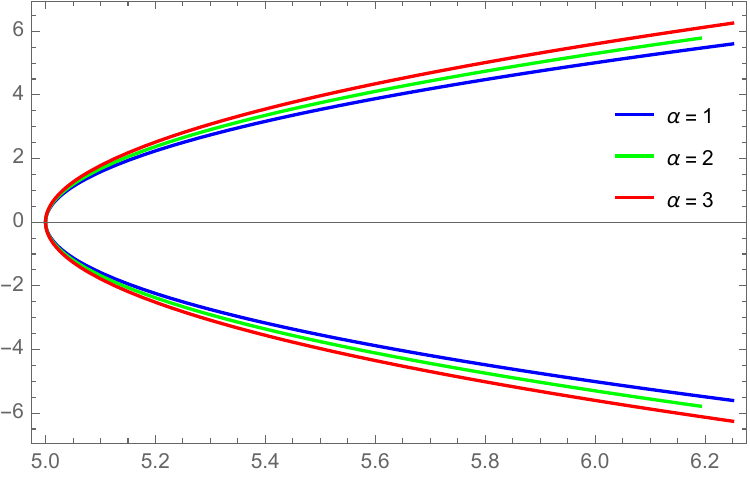}	
    \includegraphics[width=0.49\textwidth]{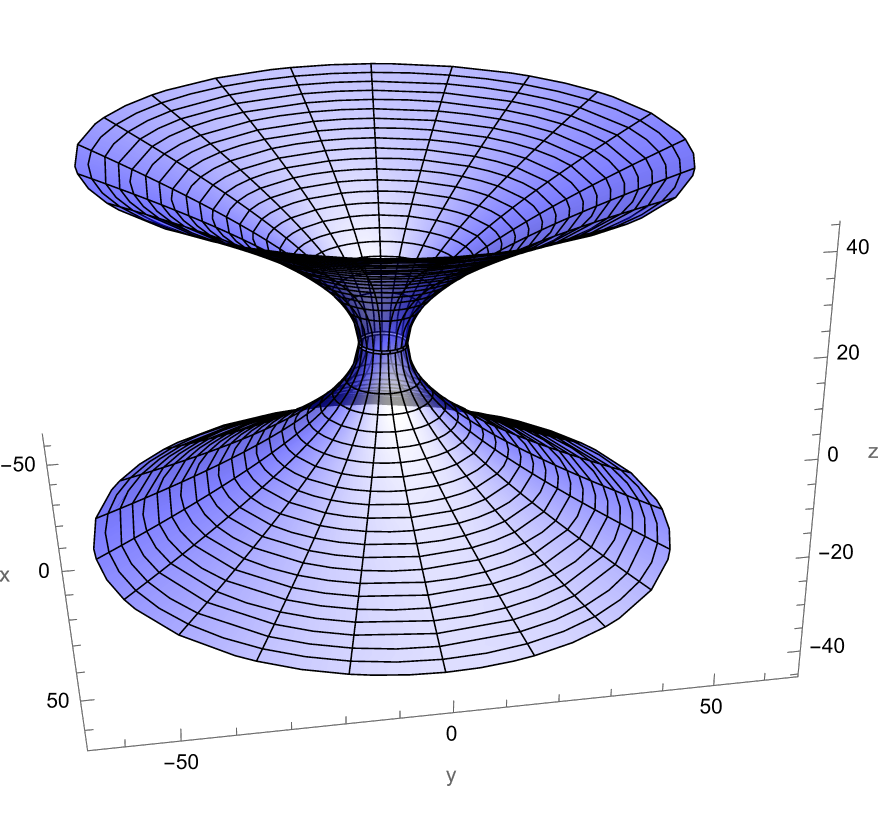}	
    \caption{Left panel: Embedding profiles $z(r)$ of the wormhole in 4D EGB gravity surrounded by a smoothed string fluid, plotted for selected values of the coupling constant \(\alpha\). The parameters are set to \(r_0 = 5\), \(a = 1\), and \(\varepsilon = 0.1\). Right panel: Three-dimensional embedding diagram of the wormhole for \(\alpha = 0.5\), using the same remaining parameters.}
    \label{wormhole_throat}
\end{figure}

The corresponding embedding diagrams, displayed in figure~\ref{wormhole_throat}, provide a direct visualization of the wormhole geometry and confirm the outward flaring of the throat.The fulfillment of the condition discussed in the previous point is illustrated by the shape of the embedding diagrams shown in figure~\ref{wormhole_throat}, for some values of $\alpha$. These diagrams generated from the mapping of the metric spatial sector in cylindrical coordinates at the equatorial plane, via
\begin{equation}
   dr^2+r^2d\phi^2+dz^2\rightarrow\frac{dr^2}{1-\frac{b(r)}{r}}+r^2d\phi^2, \nonumber
\end{equation}
provides a direct visualization of the wormhole geometry and confirm the outward flaring of the throat. Note that the greater is $\alpha$, the longer it takes for the wormhole to approach flatness. On the other hand, its curvature appears to be less pronounced at the throat, indicating a smoother transition between the regions connected by the wormhole. This feature is confirmed by the analysis of the curvature scalar made in the next section.

\subsection{Curvature regularity}

To assess the regularity of the solution, we analyze the Kretschmann scalar defined by $K \equiv R^{\mu\nu\rho\sigma} R_{\mu\nu\rho\sigma}$. This scalar provides a direct measure of the curvature intensity of spacetime, making it a valuable tool for identifying singularities, 
\bea
K &=& \frac{4 b(r)^2}{r^6} + \frac{2}{r^4}\left(b'(r)-\frac{b(r)}{r}\right)^2+8\left(1-\frac{b(r)}{r}\right)^2\frac{\Phi '(r)^2}{r^2} \nonumber\\ 
&+&\left[\left(b'(r)-\frac{b(r)}{r}\right)\frac{\Phi'(r)}{r}-2\left(1-\frac{b(r)}{r}\right)  [\Phi''(r)+\Phi'(r)^2]\right]^2.\qquad\qquad
\eea
For the zero-tidal-force wormhole, $\Phi(r)=$ const, it reduces to 
\bea
K = \frac{4 b(r)^2}{r^6} + \frac{2}{r^4}\left[b'(r)-\frac{b(r)}{r}\right]^2.
\eea
\begin{figure}[!ht]
    \includegraphics[width=0.495\textwidth]{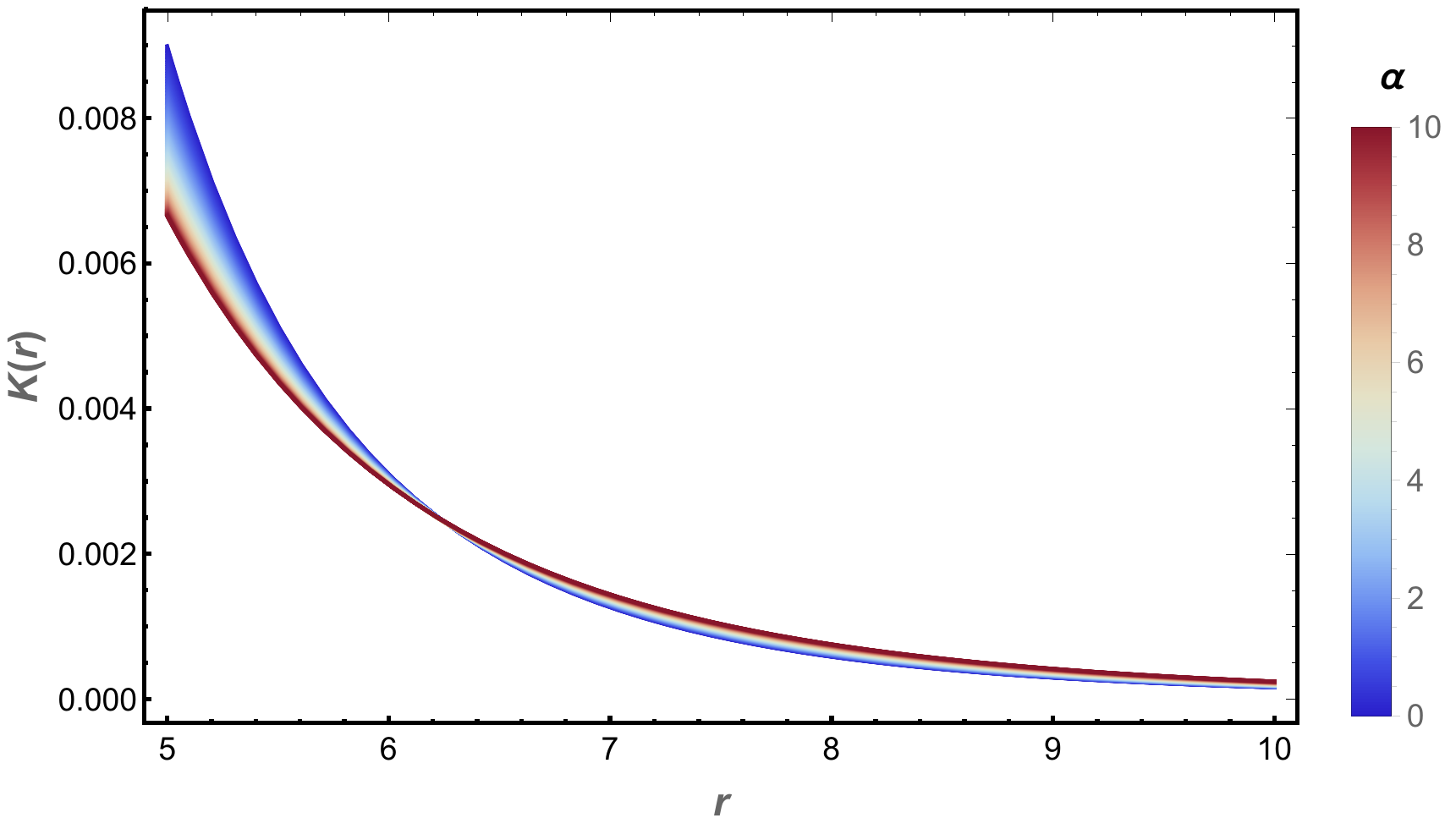}	
    \includegraphics[width=0.495\textwidth]{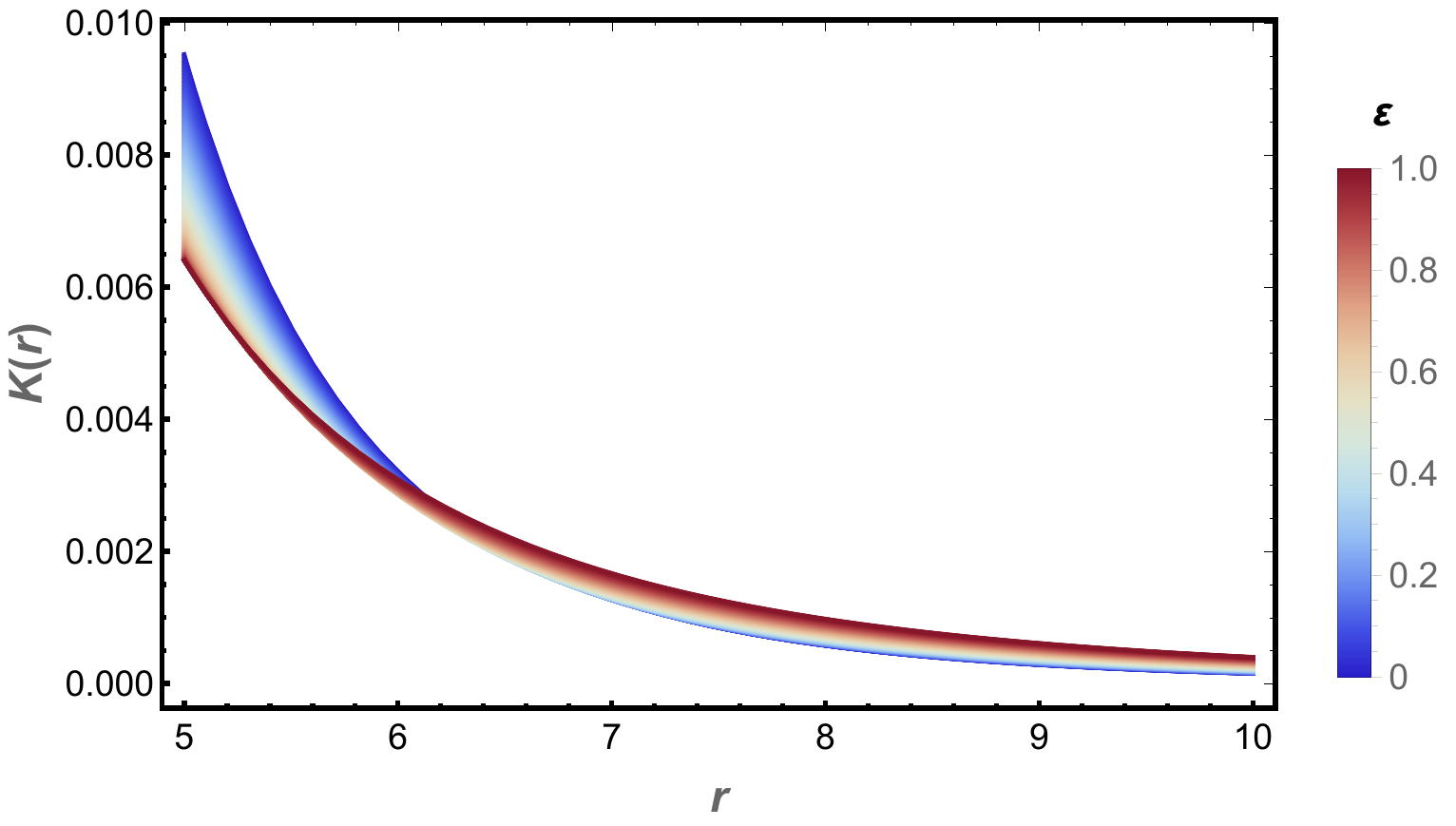}	
    \caption{Left panel: Kretschmann curvature for the wormhole in 4D EGB gravity surrounded by a smoothed string fluid, as a function of $r$. The color gradient shows the GB coupling constant in the range $\alpha \in [0,10]$ (fixed $\varepsilon=0.1$, $r_0=5$, and $a=1$). Right panel: Kretschmann curvature plotted as a color gradient for \(\varepsilon \in [0,1]\), with fixed \(\alpha = 0.1\), \(r_0 = 5\), and \(a = 1\).}
    \label{fig_Curvature_R_K}
\end{figure}
As shown in figure \ref{fig_Curvature_R_K}, the curvature remains finite throughout the spacetime, attaining its maximum near the throat and decaying smoothly at large distances. The Gauss–Bonnet coupling $\alpha$ plays a crucial role in reducing the curvature concentration near the throat, while the string fluid parameter $\varepsilon$ modulates the overall curvature scale. This interplay guarantees a regular, singularity-free geometry for all admissible parameter values. The solution's regularity persists across the full parameter space, demonstrating how modified gravity and exotic matter jointly sustain the geometry.

\subsection{Linear equations of state}

\begin{figure}[ht!]
    \centering 
    \includegraphics[width=0.49\textwidth]{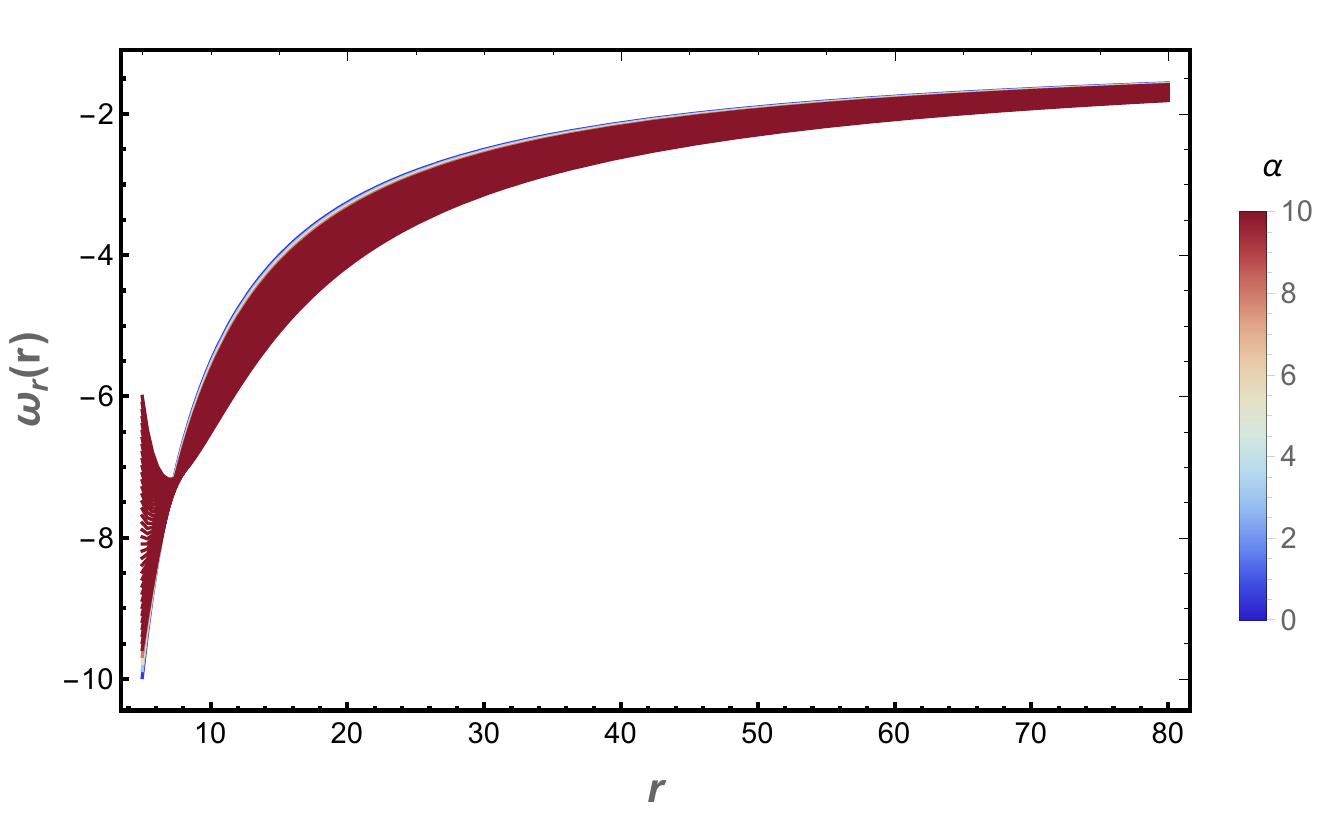}
     \includegraphics[width=0.49\textwidth]{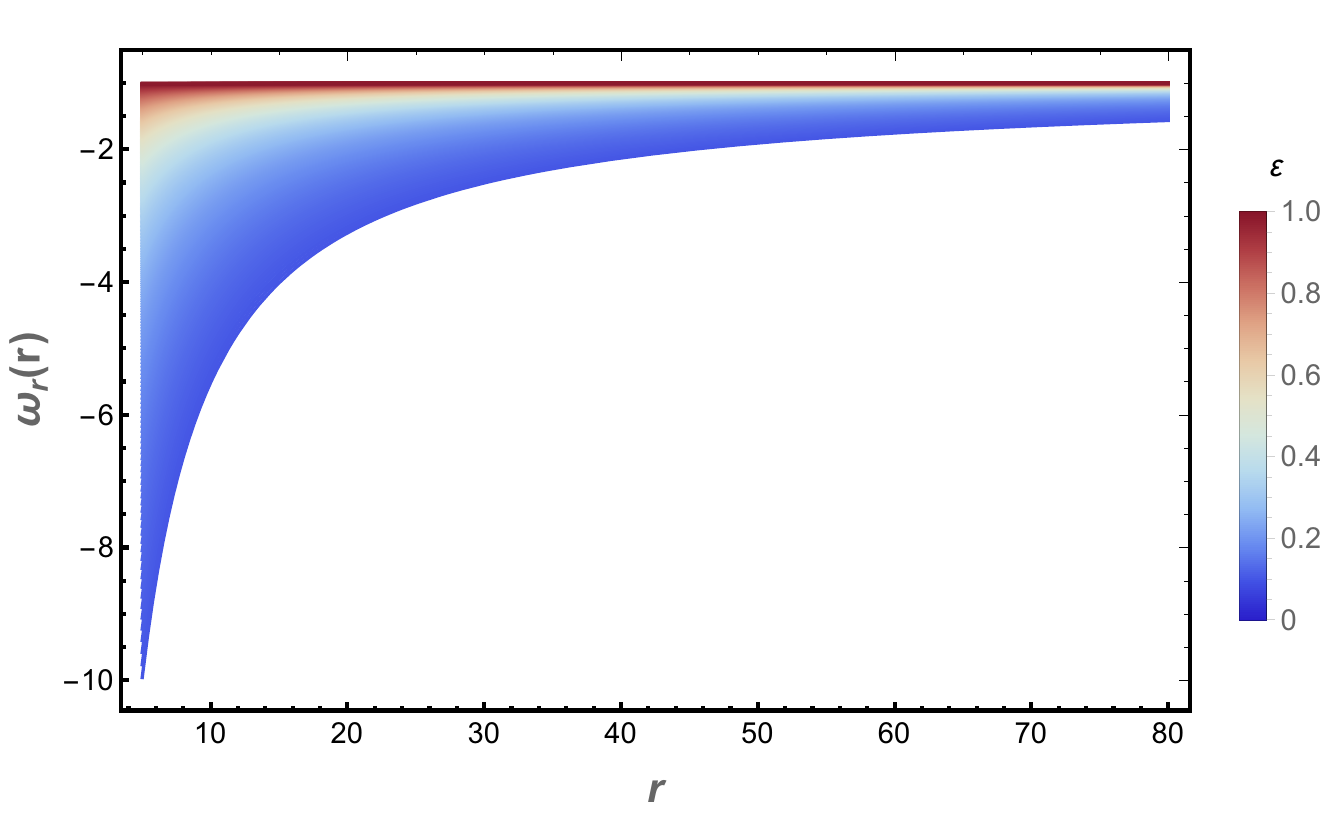}	
     \includegraphics[width=0.495\textwidth]{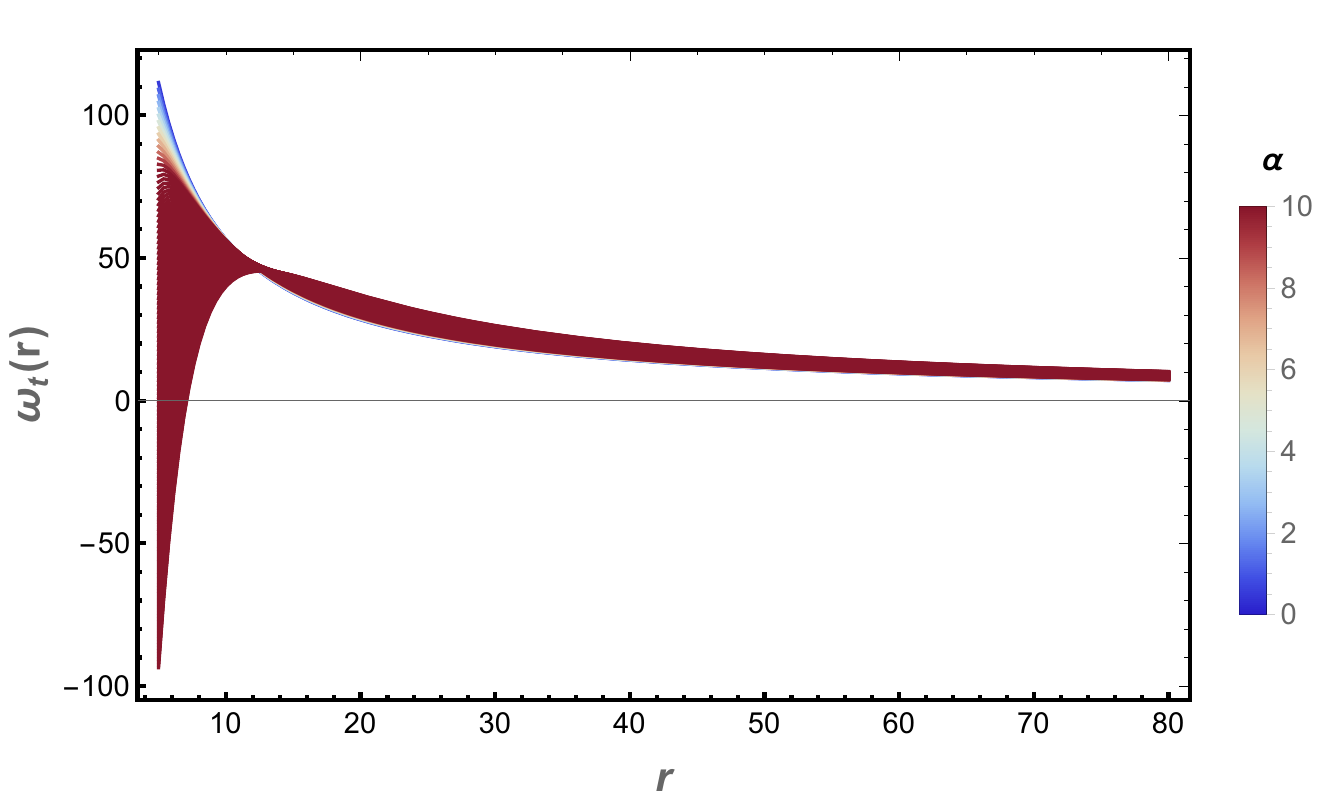}	
     \includegraphics[width=0.495\textwidth]{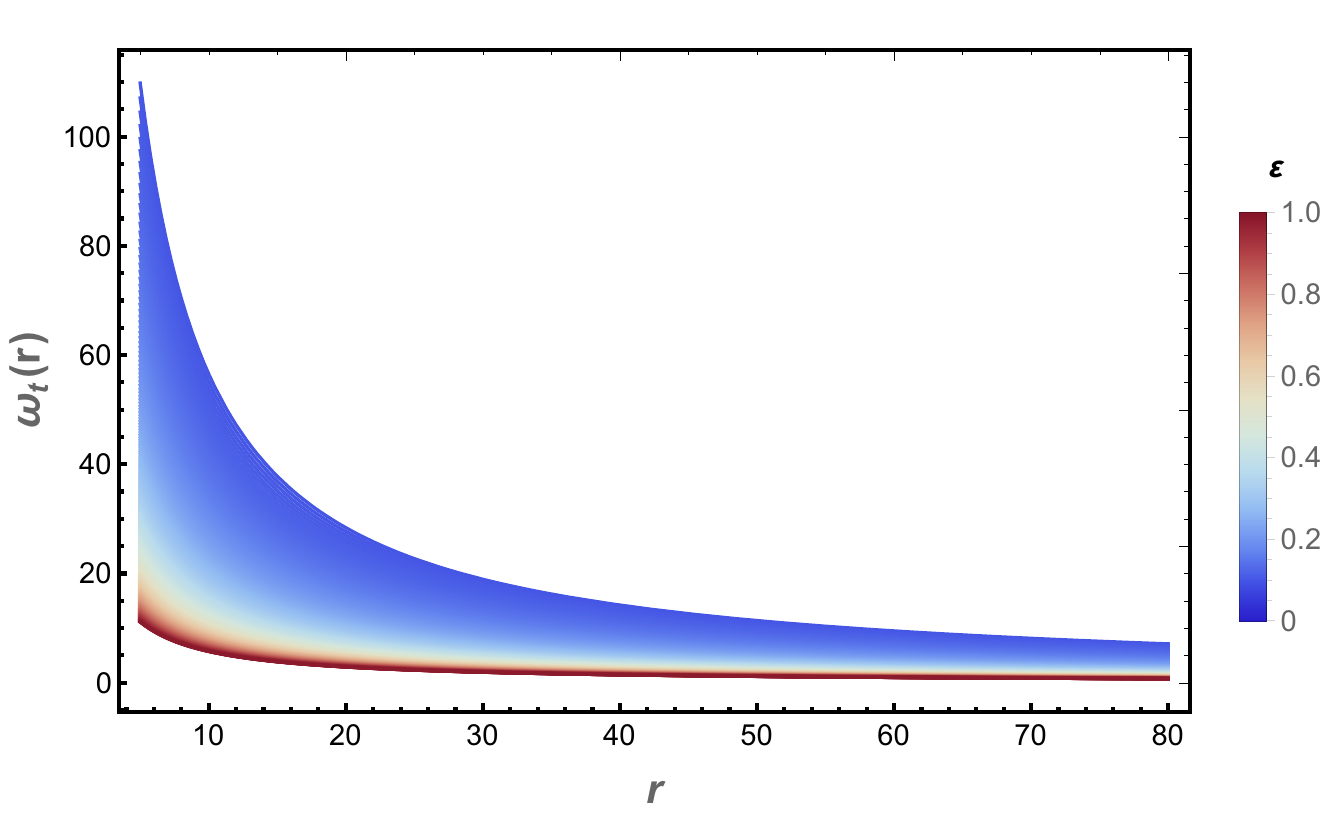}	
       \caption{Radial state parameter $\omega_r(r)$ as a function of $r$, for $\alpha \in [0,10]$ (color gradient, top left panel, fixed $\varepsilon=0.1$). On the top right panel, the color gradient shows the radial state $\omega_r(r)$ for $\varepsilon \in [0,1]$ (fixed $\alpha=0.1$). Lateral state parameter $\omega_t(r)$ as a function of the coordinate $r$, for $\alpha \in [0,10]$ (bottom left panel, fixed $\varepsilon=0.1$), and for $\varepsilon \in [0,1]$ (bottom right panel, fixed $\alpha=0.1$). In all cases, we have considered $r_0 = 5.0$ and  $a=1.0$.}
    \label{radialomega}
\end{figure}

As expected, the wormhole geometry does not satisfy the black hole EoS, $\rho = -p_r$. We therefore introduce the radial and transverse EoS parameters $\omega_r=p_r/\rho$ and $\omega_t=p_t/\rho$, where $\rho$ is given by equation~(\ref{rho}). As shown in figure \ref{radialomega}, top panels, the radial pressure exhibits a phantom-like behavior near the throat, with $\omega_r < -1$, while asymptotically approaching $-1$, recovering $p_r\approx - \rho$. Larger values of $\alpha$ significantly soften this behavior, reducing the degree of exoticity. 

The transverse EoS parameter $\omega_t(r)= p_t(r)/\rho(r)$ tends asymptotically to zero, signaling the recovery of a cloud-of-strings behavior far from the throat, where $p_t=p_{\phi}=p_{\theta}$. It represents the second EoS satisfied by the smoothed string fluid~\cite{NunesdosSantos:2025alw}. In figure~\ref{radialomega}, we show the behavior of $\omega_t$ as a function of $r$ for a continuous range of values of $\alpha$ and $\varepsilon$, displayed respectively in the bottom left and bottom right panels. As expected, $\omega_t \to 0$ asymptotically, indicating the characteristic behavior of a cloud of strings.

\subsection{Smoothed string fluid described by a Kiselev-like source}

We now consider the behavior of the linear EoS parameter identified with a Kiselev-like source, which also can be associated with an anisotropic smoothed string fluid ~\cite{NunesdosSantos:2025alw}.  
We use the fact that 
\begin{eqnarray}
  \bar{p}= \langle T^i_j \rangle =  \frac{p_r+2p_t}{3}= \omega_q \rho \,\, \quad \Longrightarrow \quad
    \omega_q = \frac{p_r+2p_t}{3\rho}. \label{EcuacionEstado1}
\end{eqnarray}
The conservation equation is such that 
\beq
    (p_r)'+ \frac{2}{r} (p_r-p_t)=0. \label{cons_eq}
\eeq 
Using \eqref{cons_eq} into \eqref{EcuacionEstado1}, we have
\begin{equation}
    \frac{p_r}{\rho} + \frac{r (p_r)'}{3 \rho} = \omega_q .
\end{equation} 
From the equation of motion, for the given value of the integration constant $\beta$ 
\bea\label{p_r}
    p_r &=& \frac{1}{8 \pi \alpha} + \frac{\alpha + r_0^2 (1 - \varepsilon) + r_0 r \varepsilon + e^{-\frac{r_0^3}{a^3}} r_0 \varepsilon (r_0 - r)}{8 r_0 \pi r^3} \nonumber \\
       &-& \frac{\sqrt{r_0 r^3 + 4 \alpha^2 + 4\alpha e^{-\frac{r_0^3}{a^3}} r_0 \varepsilon (r_0-r) + 4\alpha \left( r_0^2 (1 -\varepsilon) + r_0 r \varepsilon \right)}}{8 \sqrt{r_0} \pi r^{3/2} \alpha}.
\eea 

Additionally, a quintessence fluid is commonly described by an effective EoS of the form $\bar{p}=\omega_q\rho$, with $\omega_q\in[-1,0)$ \cite{Kiselev:2002dx}. Observations of the late–time accelerated expansion of the Universe constrain the dominant component driving this phase to lie within the interval $\omega_q\in[-1,-1/3)$. By contrast, values in the range $\omega_q\in[-1/3,0)$ are typically associated with quintessence-like matter, which has been shown to support asymptotically flat spacetimes \cite{Ghosh:2023nkr} as well as black holes exhibiting nontrivial thermodynamic critical behavior \cite{Bezerra:2019qkx}. In the limiting case $\omega_q=-1$, the fluid mimics a positive cosmological constant and, within the Kiselev framework \cite{Kiselev:2002dx}, gives rise to a repulsive potential analogous to that of de Sitter spacetime, thus corresponding to a de Sitter–like EoS.

On the other hand, the asymptotic limit $\omega_q\rightarrow -1/3$ is characteristic of a cloud of cosmic strings, providing a natural string–fluid interpretation of the source. Taken together, these two limits illustrate the versatility of the smoothed string fluid, which is able to interpolate between a de Sitter–like spacetime near the throat and a cosmic–string–like behavior at large distances within a single unified framework.

\begin{figure}[hb!]
       \includegraphics[width=0.455\textwidth]{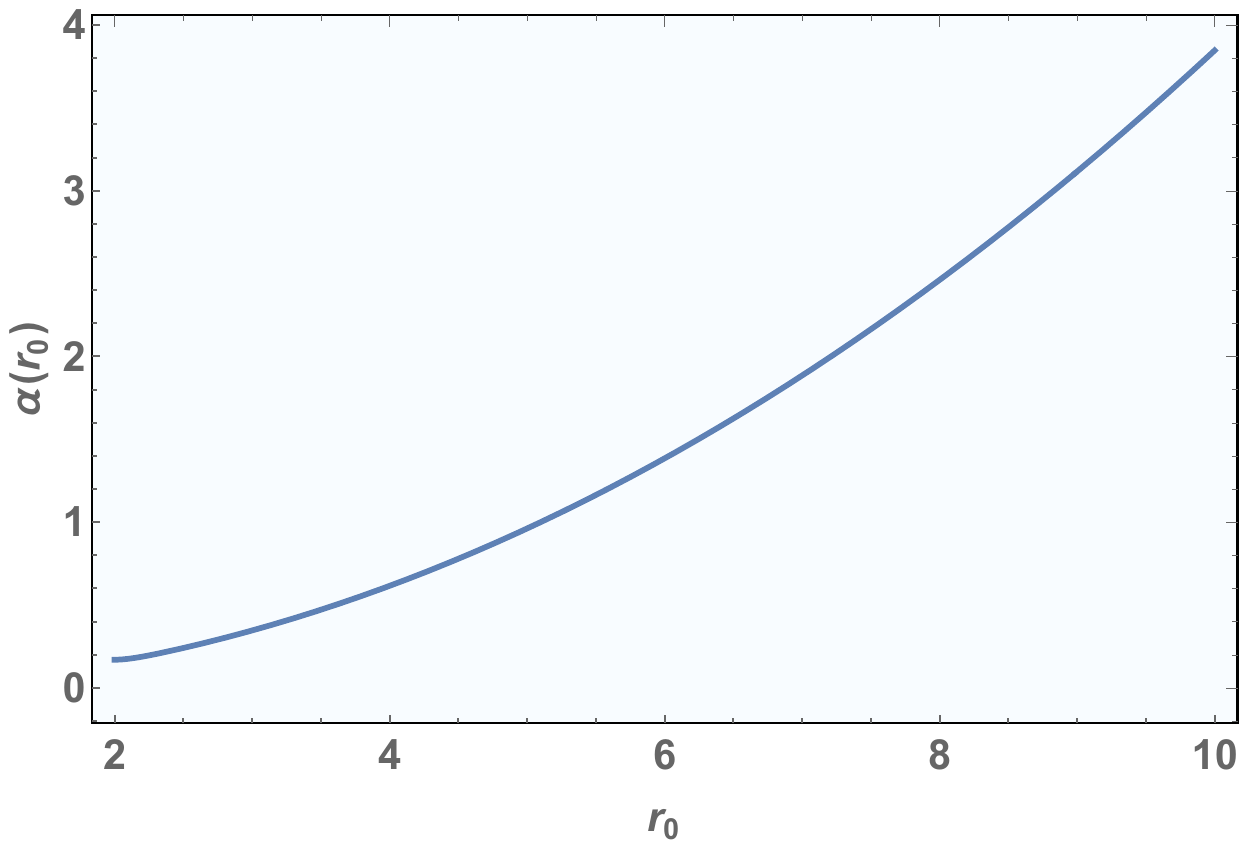}
         \includegraphics[width=0.535\textwidth]{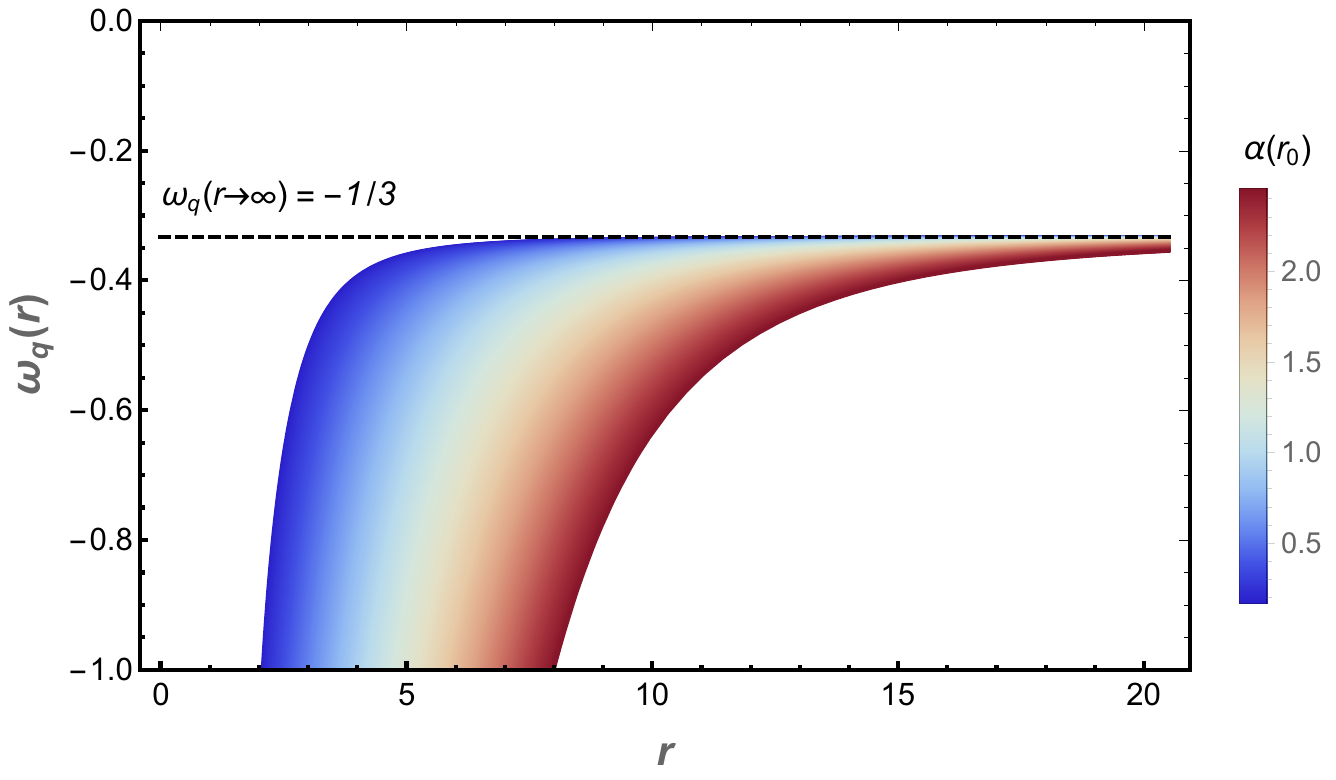}
    \caption{Left panel: Values of $r_0$ and $\alpha(r_0)$ in the parameter space such that the EoS near the throat resembles a de Sitter EoS, \textit{i.e.}, $\omega_q(r_0) = -1$. Right panel: Color gradient shows $\omega_q(r)$ for $r_0 \in[2,8]$. In both cases, we used fixed values of $\varepsilon = 0.1$ and $a = 1$.}
 \label{figEjemplosOmega}
\end{figure}

The condition $\omega_q(r_0)=-1$ imposes a constraint between the wormhole throat radius $r_0$, the string fluid density $\varepsilon$, and the GB coupling $\alpha$. For fixed $\varepsilon$ and length scale $a$, we found that
\bea \label{alpha_r0}
\alpha(r_0)= r_0^2 \left[\frac{\varepsilon (e^{r_0^3/a^3}-1)+3 (\varepsilon+1) r_0^3/a^3}{e^{r_0^3/a^3}(3-4 \varepsilon)+4\varepsilon -12 (\varepsilon+1)r_0^3/a^3}\right].
\eea
This relation reveals how $\alpha$ must scale with $r_0$ in order to maintain $\omega_q(r_0)=-1$. Figure \ref{figEjemplosOmega} (left panel) illustrates this for $\varepsilon=0.1$ and $a=1$. The right panel of figure \ref{figEjemplosOmega} confirms that this parameter choice indeed yields $\omega_q(r_0)=-1$, while asymptotically approaching $\omega_q\rightarrow -1/3$ (cosmic string behavior, dashed line). Notably, this asymptotically approaching limit aligns with observational constrains on late-time cosmic acceleration \cite{Kiselev:2002dx, Vilenkin:1981kz, Nojiri2017}. 

Thus, those figures illustrate an example within the parameter space in which the EoS satisfies $\omega_q \in [-1, -1/3]$, \textit{i.e.}, consistent with the constraints associated with the late-time cosmic acceleration of the universe, while the asymptote of the EoS ($\omega_q \to -1/3$) resembles the behavior of a cosmic string \cite{Vilenkin:1981kz,Nemiroff:2007xs}. Ref. \cite{Fernandez-Nunez:2016urh} states that a cosmic string can be understood as a long-lived, topologically stable structure that may have been formed during phase transitions in the early Universe. The spacetime around a straight cosmic string is locally flat but globally exhibits a conical topology. The authors assert that the most evident way to detect cosmic strings is through gravitational lensing.
\newline

\section{Null Energy Condition and Quantity of Exotic Matter}

The null energy condition (NEC) plays a central role in the physics of traversable wormholes. In General Relativity, the flaring–out condition at the throat necessarily requires the violation of the NEC, implying the presence of exotic matter. In modified theories of gravity, however, higher–curvature terms can effectively contribute to the stress–energy balance, partially replacing the role usually played by exotic sources. In this section, we analyze how the EGB coupling $\alpha$ and the smoothed string fluid parameter $\varepsilon$ affect the fulfillment of the NEC and the total amount of exotic matter required to support the wormhole geometry.

For a static, spherically symmetric spacetime, the NEC is expressed by the following conditions:
\beq
    \rho + p_r \geq 0, \quad \rho + p_t \geq 0,
\eeq
where $\rho$, $p_r$, and $p_t$ are the energy density, radial pressure, and tangential pressure, respectively. These quantities are evaluated from the field equations for the wormhole metric introduced in Section \ref{wsolutions}. 

\begin{figure}[ht!]
    \centering 
    \includegraphics[width=0.49\textwidth]{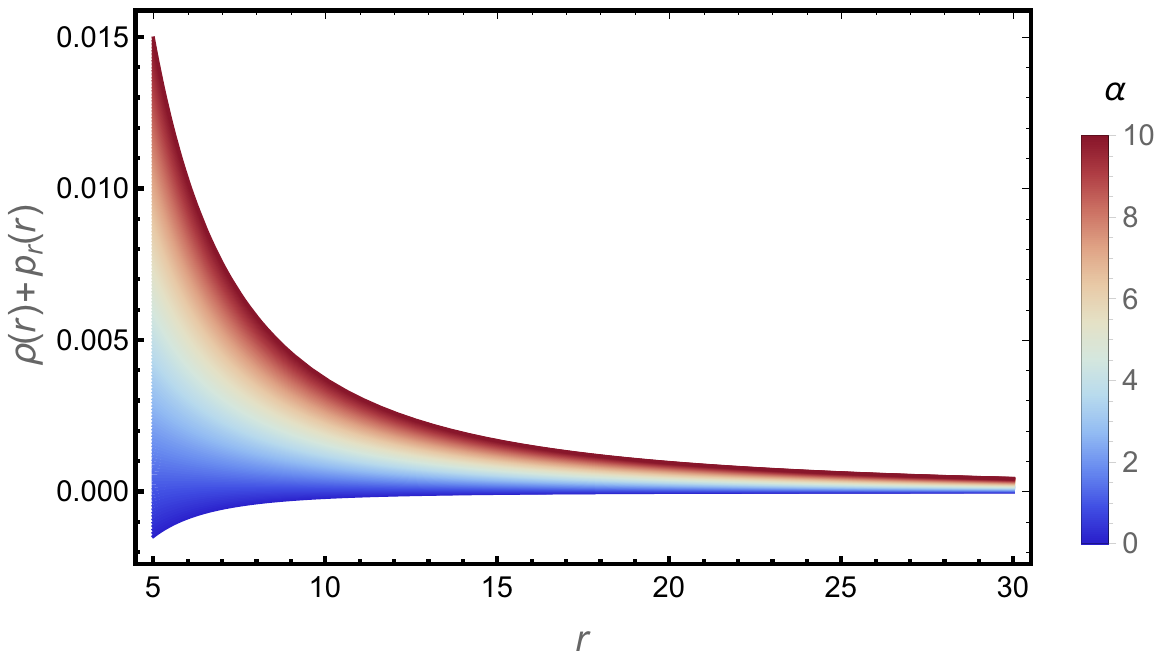}
    \includegraphics[width=0.49\textwidth]{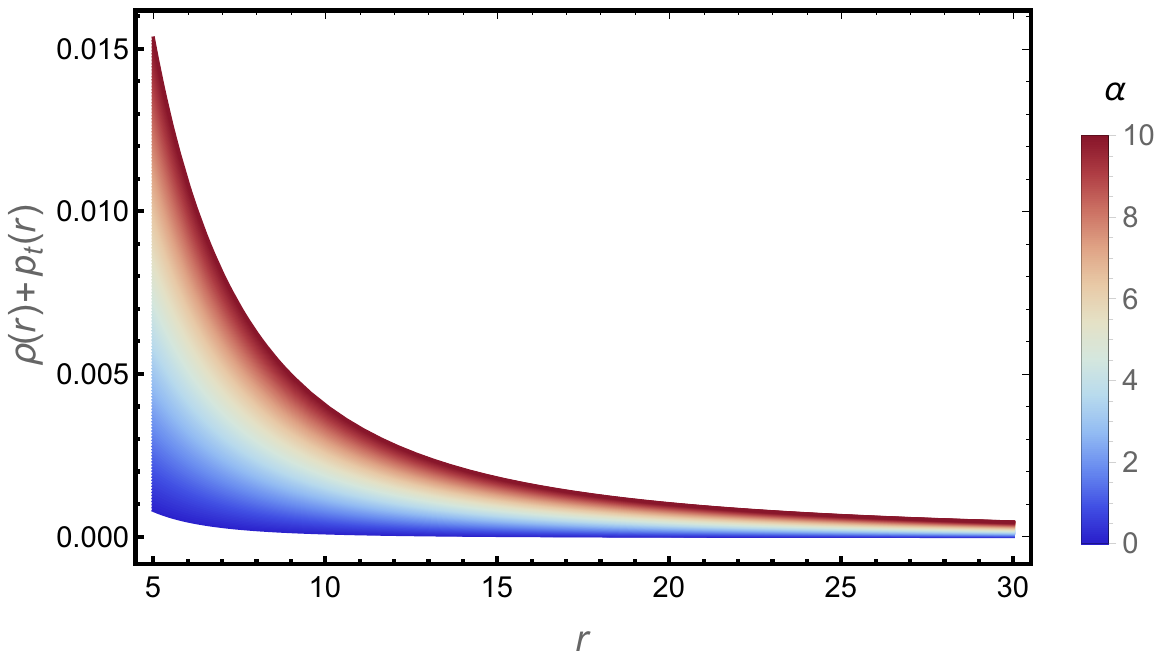}
    \includegraphics[width=0.49\textwidth]{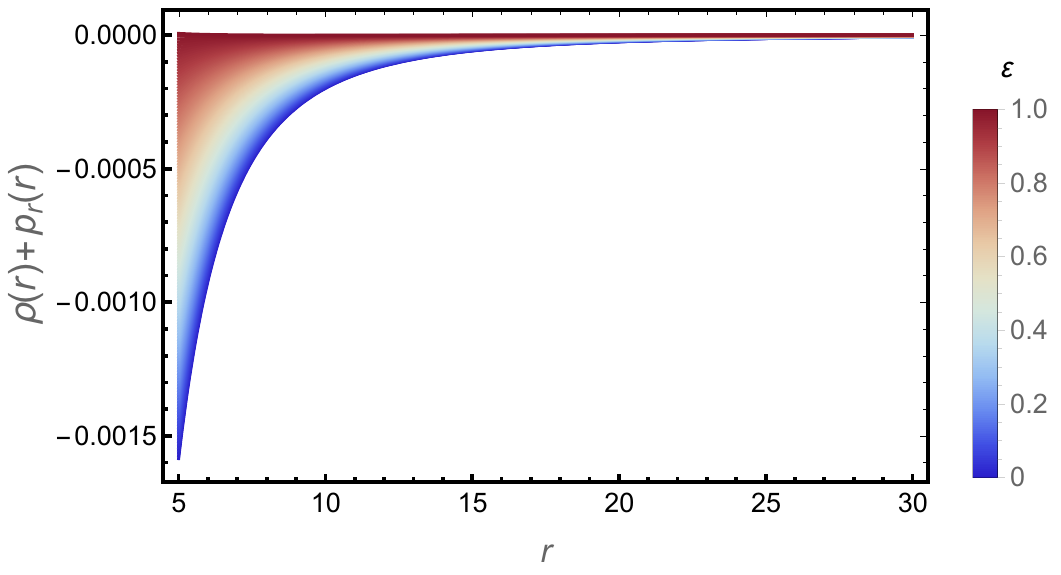}
    \includegraphics[width=0.49\textwidth]{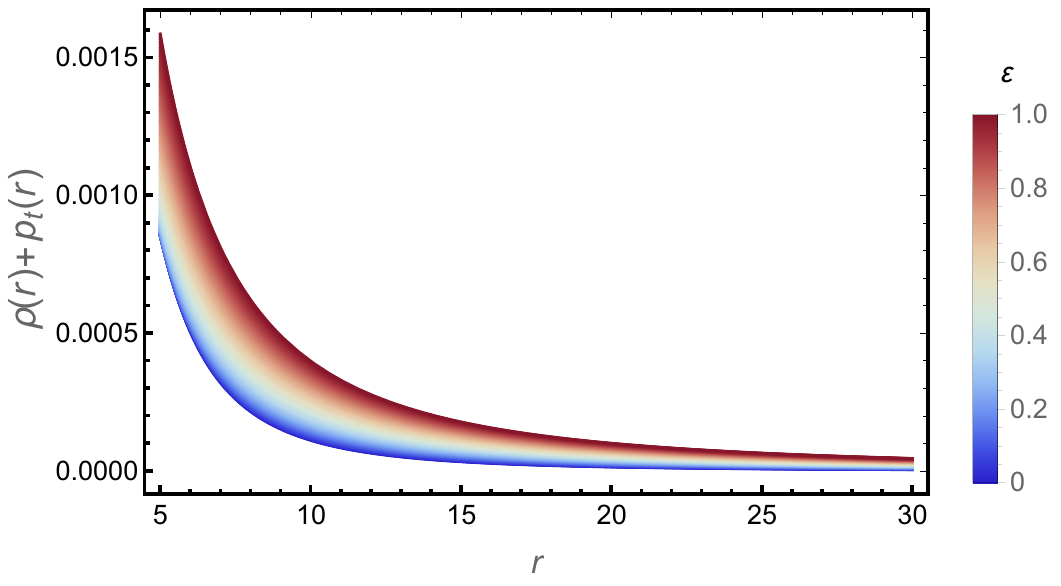}
     \caption{NEC profiles for the wormhole in EGB gravity supported by a smoothed string fluid. The top panels display $\rho+p_r$ (left) and $\rho+p_t$ (right) for several values of the Gauss–Bonnet coupling $\alpha$, with $r_0=5$, $a=1$, and $\varepsilon=0.1$. The bottom panels show $\rho+p_r$ (left) and $\rho+p_t$ (right) for a fixed coupling $\alpha=0.1$ and a continuous range of the smoothed string fluid parameter $\varepsilon\in(0,10)$.}
    \label{fig7}
\end{figure}

The radial and transverse NEC combinations are shown in figure \ref{fig7} for representative values of the parameters. The top panels display $\rho+p_r$ and $\rho + p_t$ for a range of values of the Gauss–Bonnet coupling $\alpha$, while the bottom panels show their behavior for a fixed coupling $\alpha = 0.1$. The figures indicate that any violation of the NEC is confined to a narrow region around the throat and that increasing  $\alpha$ significantly weakens both the magnitude and the spatial extent of this violation. This behavior reflects the fact that, in EGB gravity, part of the effective stress–energy required to sustain the flaring–out condition is provided by the higher–curvature sector rather than by the physical matter fields alone. 

To quantify the total amount of exotic matter necessary to sustain the wormhole, we adopt the Volume Integral Quantifier (VIQ) introduced in Ref.~\cite{Nandi:2004ku}, defined by
\begin{equation}\label{viq}
\mathcal{I}_v = \oint 4\pi r^2 (\rho + p_r)\,dr = 2\int_{r_0}^r 4\pi x^2 (\rho + p_r)\,dx.
\end{equation}
which measures the integrated violation of the NEC over the spatial volume. A negative value of $\mathcal{I}_v$ indicates the presence of exotic matter, whereas a vanishing or positive value corresponds to a configuration that can be supported without NEC–violating sources. 
 
Here, we are particularly interested in evaluating this quantity near the wormhole throat. Although wormhole geometries are generally associated with violations of the null energy condition (NEC), our analysis reveals that for sufficiently high values of the Einstein-Gauss-Bonnet coupling and low values of the smoothed string fluid parameter, the NEC may remain satisfied in the vicinity of the throat. In such cases, the integral $\mathcal{I}_v$ can approach zero or even become positive, suggesting that the wormhole could be supported without the need for exotic matter in this regime. 

Using equations (\ref{eqrho}) and (\ref{radpress}), and expanding the integrand of equation \eqref{viq} to first order around $r = r_0 $, we find that the VIQ is approximately given by
\begin{equation}
  \mathcal{I}_v \approx  \left[-1 + \frac{3r_0 e^{-\frac{r_0^3}{a^3}}}{8 a^3 \pi} + \frac{\alpha}{r_0^2} 
+ \frac{1}{8\pi r_0^2}\left( 1 - {e^{-\frac{r_0^3}{a^3}}} 
+ \frac{3r_0^3}{a^3}e^{-\frac{r_0^3}{a^3}} \right) \varepsilon\right](r - r_0).
\end{equation}
This expression shows that, for fixed $r_0$, $a$, and $\varepsilon$, the VIQ can be made arbitrarily small by increasing the GB coupling constant $\alpha$. In contrast to General Relativity, where the VIQ is necessarily negative and sizable, the EGB corrections drive $\mathcal{I}_v$ toward zero, indicating a progressive suppression of the total amount of exotic matter, that require violation of the NEC, even in models with frozen scalar fields \cite{Elizalde2024}. 

Therefore, although the NEC may still be locally violated near the throat, the higher–curvature contributions substantially reduce the energetic cost of maintaining the wormhole. In the strong–coupling regime of EGB gravity and for sufficiently small values of the smoothed string fluid parameter $\varepsilon$, the integrated NEC violation can become negligible, allowing the wormhole geometry to be sustained with little or no exotic matter. This result is consistent with the behavior of the radial equation–of–state parameter discussed throughout the previous section and confirms the role of higher–curvature effects in softening the classical energy–condition constraints.

\section{Complexity Factor}

The internal structure of a self–gravitating system can be characterized not only by its energy conditions, but also by the degree of anisotropy and inhomogeneity of its matter distribution. In this context, the complexity factor introduced by Herrera \cite{Herrera:2018bww} provides a useful scalar measure of the the degree of structural complexity required to sustain a given spacetime geometry. In wormhole configurations, large anisotropies and strong density gradients are typically associated with the presence of exotic matter, making the complexity factor an efficient mathematical tool for diagnosing wormhole solutions.

For the present EGB wormhole supported by a smoothed string fluid, the complexity factor is given by
\begin{equation}
    Y_{TF}=8\pi(p_r-p_t)-\frac{4\pi}{r^3}\int_{r_0}^{r}\!\! u^3\frac{d\rho(u)}{du}du.
\end{equation}

\begin{figure}[!ht]
    \centering 
    \includegraphics[width=0.49\textwidth]{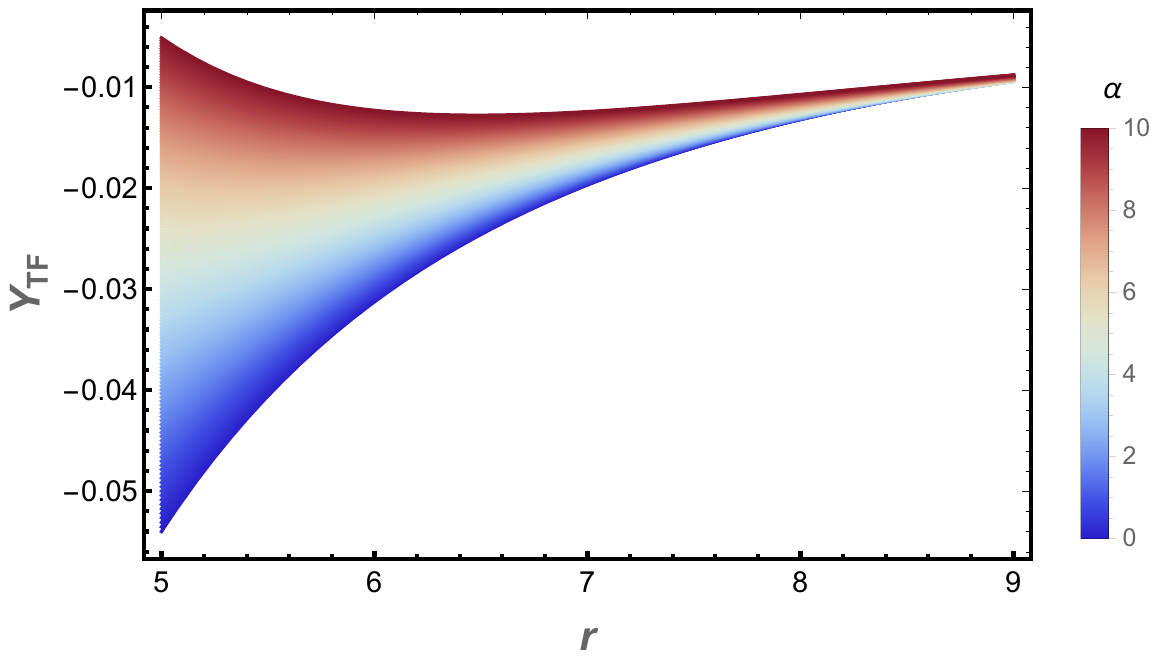}
    \includegraphics[width=0.49\textwidth]{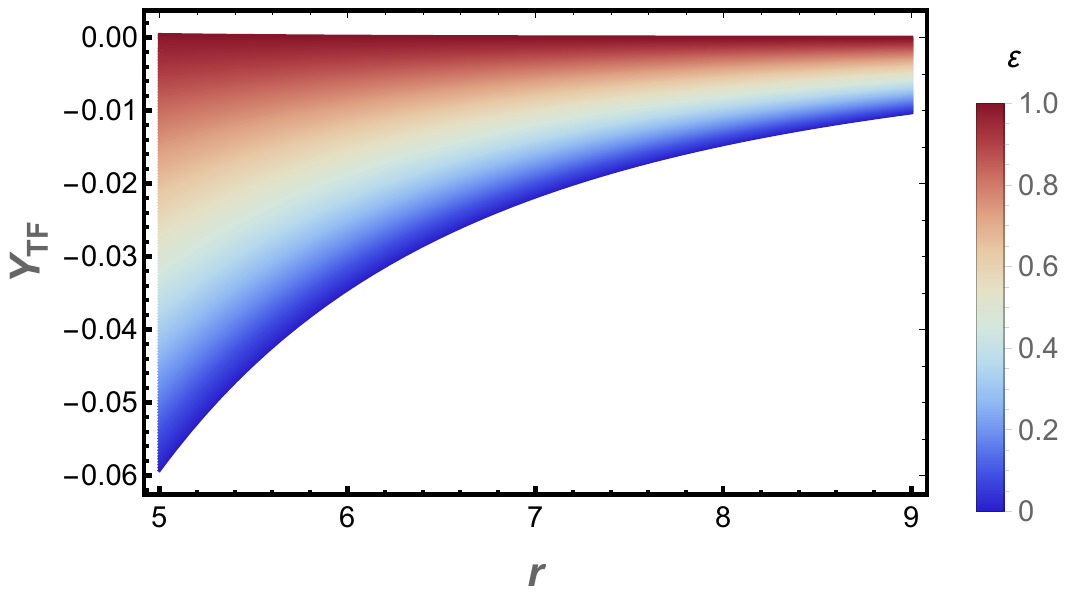}
     \caption{Radial profiles of the complexity factor $Y_{TF}$ for the wormhole in EGB gravity supported by a smoothed string fluid. Left panel: dependence on the Gauss–Bonnet coupling $\alpha$ for a fixed string fluid parameter $\varepsilon = 0.1$. Right panel: dependence on the smoothed string fluid parameter $\varepsilon$ for a fixed coupling $\alpha = 0.1$. All curves are shown for $r_0 = 5$ and  $a = 1$. The decrease of $|Y_{TF}|$ with increasing $\alpha$ and $\varepsilon$ indicates a progressive reduction of the structural complexity of the wormhole configuration.}
    \label{complex}
\end{figure}
The radial behavior of $Y_{TF}$ is displayed in figure~\ref{complex}. The left panel shows the dependence on the Gauss-Bonnet coupling $\alpha$ for a fixed value of the string fluid parameter $\varepsilon = 0.1$, while the right panel illustrates the effect of varying $\varepsilon$ for a fixed coupling $\alpha = 0.1$. 

The results display a systematic dependence, namely that increasing the Gauss–Bonnet coupling leads to a systematic reduction in the magnitude of the complexity factor throughout the spacetime, particularly near the wormhole throat. This indicates that higher–curvature corrections effectively simplify the internal structure of the wormhole, reducing the degree of anisotropy and inhomogeneity required to support the flaring–out geometry. In this sense, the Gauss–Bonnet term acts as a geometric contribution that partially replaces the role of exotic matter.

A similar behavior is observed when the smoothed string fluid parameter $\varepsilon$ is increased. As shown in the right panel of figure \ref{complex}, larger values of $\varepsilon$ lead to smaller values of $Y_{TF}$, driving the configuration toward the limit $Y{TF} \rightarrow 0$, which corresponds to a non–locally constrained equation of state. This indicates that the anisotropic stresses associated with the string fluid contribute to stabilizing the wormhole with reduced structural complexity.

Taken together, these results provide an independent confirmation that the combined effect of the smoothed string fluid and the higher–curvature corrections reduces both the integrated NEC violation and the complexity factor as $\alpha$ and $\varepsilon$ increase, allowing the wormhole to be supported with a progressively smaller amount of exotic matter and structural anisotropy.

\section{Conclusion}
\label{sec:conclusion}

Our analysis of traversable wormholes in four–dimensional Einstein–Gauss–Bonnet gravity supported by a smoothed string fluid has revealed a consistent and physically transparent mechanism for reducing the need for exotic matter. The wormhole geometry satisfies all standard traversability requirements: the shape function $b(r)$ defines a regular throat, while the constant redshift function $\Phi(r)$ guarantees a zero–tidal–force configuration free of horizons. Regularity is further ensured by the smoothed string fluid profile, originally introduced to regularize black–hole interiors \cite{NunesdosSantos:2025alw}, which here also leads to a completely nonsingular wormhole spacetime, as confirmed by the bounded behavior of the curvature invariants.

The matter sector admits a dual interpretation: microscopically as a string fluid with a radially varying equation of state, and macroscopically as an effective Kiselev–type anisotropic source. This mapping clarifies the physical content of the solution, showing that the fluid interpolates between a de Sitter–like core, characterized by $\omega_r \rightarrow -1$, and a cosmic–string–dominated exterior, associated with the asymptotic limit $\omega_t \rightarrow -1/3$. The Gauss–Bonnet coupling $\alpha$ controls how smoothly this transition occurs and how strongly curvature is suppressed near the throat.

The embedding diagrams and curvature profiles demonstrate that increasing GB coupling $\alpha$ systematically reduces the curvature at the throat and drives the geometry toward a smoother, nearly flat configuration. This geometric regularization is mirrored by the behavior of the null energy condition (NEC). While classical general relativity requires NEC violation to sustain wormholes, in the EGB framework the NEC can be satisfied in extended regions of spacetime. For sufficiently strong Gauss–Bonnet coupling and moderate values of the smoothed string parameter $\varepsilon$, both $\rho+p_r$ and $\rho+p_t$ remain non–negative at and outside the throat, indicating that higher–curvature corrections effectively soften the energy–condition violations.

This conclusion is quantitatively supported by the volume integral quantifier and the complexity factor. The VIQ shows that the integrated amount of NEC–violating matter near the throat decreases with increasing $\alpha$, while the complexity factor $Y_{TF}$  exhibits a systematic reduction as either $\alpha$ or $\varepsilon$ grows. In particular, the smoothed string fluid drives the configuration toward the low–complexity limit, with $Y_{TF} \rightarrow 0$ as $\varepsilon$ increases, signaling a progressive suppression of anisotropy and density inhomogeneity. These two diagnostics together demonstrate that the same physical mechanism responsible for weakening the NEC violation also reduces the internal structural complexity of the wormhole.

The analysis of the radial and transverse equations of state further supports this picture. For small $\alpha$ and $\varepsilon \rightarrow 1$, the matter content approaches that of a pressureless cloud of cosmic strings, with vanishing transverse pressure, while larger values of $\alpha$ allow this string–dominated configuration to be smoothly embedded within a regular, traversable geometry. The combined action of higher–curvature corrections and the smoothed string fluid therefore enables wormhole solutions that are both geometrically regular and energetically less exotic.

In conclusion, our results show that Einstein–Gauss–Bonnet gravity coupled to a string fluid with a variable equation of state provides a natural framework in which traversable wormholes can be sustained with a reduced amount of exotic matter and lower structural complexity. This unifies regular black–hole cores, Kiselev–type anisotropic fluids, and traversable wormholes within a single coherent model. Future work should address linear stability, perturbative dynamics, and potential observational imprints, as well as extensions to rotating and dynamical configurations. Finally, this work opens promising avenues for understanding how string-inspired gravity modifications facilitate traversable wormhole solutions with reduced exotic matter requirements.

\begin{center}
\rule{0.5\textwidth}{0.4pt}
\end{center}

\section*{Data availability statement} 
All data that support the findings of this study are included within the article (and any supplementary
files).

\section*{Acknowledgment} This work was partially supported by Conselho Nacional de Desenvolvimento Científico e Tecnológico (CNPq) under Grants 301122/2025-3 (CRM).  LCNS would like to thank FAPESC for financial support under Grant No. 735/2024. The authors thank colleagues at Universidade Estadual do Ceará for valuable discussions.

\section*{References} 
\bibliographystyle{iopart-num}
\bibliography{ref} 
\end{document}